\documentclass[12pt]{iopart}
\usepackage{iopams}
\usepackage{graphicx}
\usepackage{bm}

\def\be{\begin{eqnarray}}
\def\ee{\end{eqnarray}}
\def\ba{\begin{array}}
\def\ea{\end{array}}
\def\nn{\nonumber}
\begin{document}


\article[Coupling asymmetries]{Quantum phases: 50 years of the Aharonov-Bohm Effect and 25 years of the Berry phase}{Investigation of the coupling asymmetries at double-slit interference experiments}

\author{\center{A. I. Mese$^1$, A. Bilekkaya$^1$, S. Arslan$^2$, S. Aktas$^1$} \\ \center{and} \\ \center{A. Siddiki$^{2,3}$}}

\address{$^1$ Trakya University, Department of Physics, 22030 Edirne, Turkey}
\address{$^2$ Department of Physics, Arnold Sommerfeld Center for Theoretical Physics, Ludwig-Maximilans-Universitat, Theresienstrasse 37, 80333 Munich, Germany}
\address{$^3$ Istanbul University, Faculty of Sciences, Physics Department, Vezneciler-Istanbul 34134, Turkey}

\begin{abstract}
Double-slit experiments inferring the phase and the amplitude of
the transmission coefficient performed at quantum dots (QD), in
the Coulomb blockade regime, present anomalies at the phase
changes depending on the number of electrons confined. This phase
change cannot be explained if one neglects the electron-electron
interactions. Here, we present our numerical results, which
simulate the real sample geometry by solving the Poisson equation
in 3D. The screened potential profile is used to obtain energy
eigenstates and eigenvalues of the QD. We find that, certain
energy levels are coupled to the leads stronger compared to
others. Our results give strong support to the phenomenological
models in the literature describing the charging of a QD and the
abrupt phase changes.
\end{abstract}
%

\section{Introduction}
One of the most interesting experiments in the history of physics
is the double-slit experiments, which infers to the quantum
mechanical nature of the particles. The technological developments
in producing low dimensional high mobility charge carrier systems,
enabled experimentalists to re-do the double-slit experiments
considering nanostructures. In the experiments performed at
cryogenic temperatures and considering a two dimensional electron
system (2DES), the phase and the transmission amplitude were
measured
simultaneously~\cite{Yacoby95:abinter1,Heiblum05:abinter}. The
findings of these and consequent experiments activated a huge
number of theoreticians to understand the physics underlying the
abrupt phase
changes~\cite{Hacken95:phase,Imry00:towards,Hacken2001:all,Theresa07:lapses,Theresa07:crossover,Imry07:lapses},
for a comprehensive review we suggest the reader to check
especially Ref.~\cite{Hacken2001:all} and the references given
thereby. In particular, G. Hackenbroich \textit{et. al}
investigated the effect of shape deformation of a parabolic
quantum dot (QD), in the absence of Coulomb interaction, and
showed that the degeneracy due to the symmetry of the QD is
lifted, however, for the deformed QD it is still possible to
obtain density of states in broad energy intervals, which are
large compared to the single particle level spacings $\delta$. The
interplay between the level width $\Gamma$ and $\delta$ is used to
give an explanation to the observed phase
anomalies~\cite{Theresa07:lapses,Imry07:lapses}. On the other
hand, the effect of interactions was included by M. Stopa by
solving the related Poisson and Schr\"odinger equations
self-consistently within a Hartree-Fock type mean field
approximation, however, its influence on the phase was left
unresolved. We should also note that, in these calculations a
rather simplified QD geometry was investigated compared to the
experiments.

This work aims to provide numerical support to the theories which
rely on the formation of a wide state at certain QD geometries. We obtain the potential profile of the real samples by
solving the Poisson equation in 3D using fast Fourier
transformation,
iteratively~\cite{Weichselbaum03:056707,Sefa08:prb}. In our
calculations, we consider the sample geometry presented in
Ref.~\cite{Heiblum05:abinter}. The next step is to obtain the
energy eigenstates and values for the calculated effective
potential. We solve the Schr\"odinger equation by diagonalizing
the single-particle Hamiltonian implementing the finite difference
techniques.
%
\section{Theory}
Here, we investigate the single particle
eigen-energies and eigenfunctions of the reduced 2D Hamiltonian
\be H=\frac{\bf{p}^2}{2m^*}+V(x,y), \ee where $\bf{p}$ is the
momentum operator in 2D, $m^*$ is the effective mass ($=0.067m_e$
in GaAs) and $V(x,y)=V_G(x,y)+V_D(x,y)+V_l(x,y)+V_H(x,y)$ is the
mean field potential composed of gates, leads, donors and Hartree
terms, respectively.
\newline
In physics, WKB approximation is one of the most frequently applied approximation to solve Schrdinger's equation.
The transmission amplitude $W_n(a,b)$ is calculated at the
barrier along the classical turning points (a,b), via \be
W_n(a,b)(E)=\frac{e^{\xi(E)}}{1+\frac{1}{4}e^{\xi(E)}},\nn \\
{\xi(E)}=-2\int_{a}^{b}dx\sqrt{\frac{2m}{\hbar^2}(V(x)-E)}.\ee
It is known that the transmission amplitude depend
almost linearly to the energy of the incoming state, assuming
plane waves and within the WKB approximation~\cite{Stopa96:selfcTF}, which we utilize likewise in the following to calculate transport through the QD.

We proceed our work by considering the real
geometry and the potential profile calculated within the
self-consistent Thomas-Fermi-Poisson (TFP) theory. In the
following section we first discuss the limitations of such a mean
field approximation and compare our method with the existing
calculation schemes in the high electron occupation regime, \emph{i.e.}
$N\gtrsim100$. We show that, the single particle energy states and
energies can be well described in this regime considering TFP
theory.\begin{figure}{\centering
\includegraphics[width=.7\linewidth]{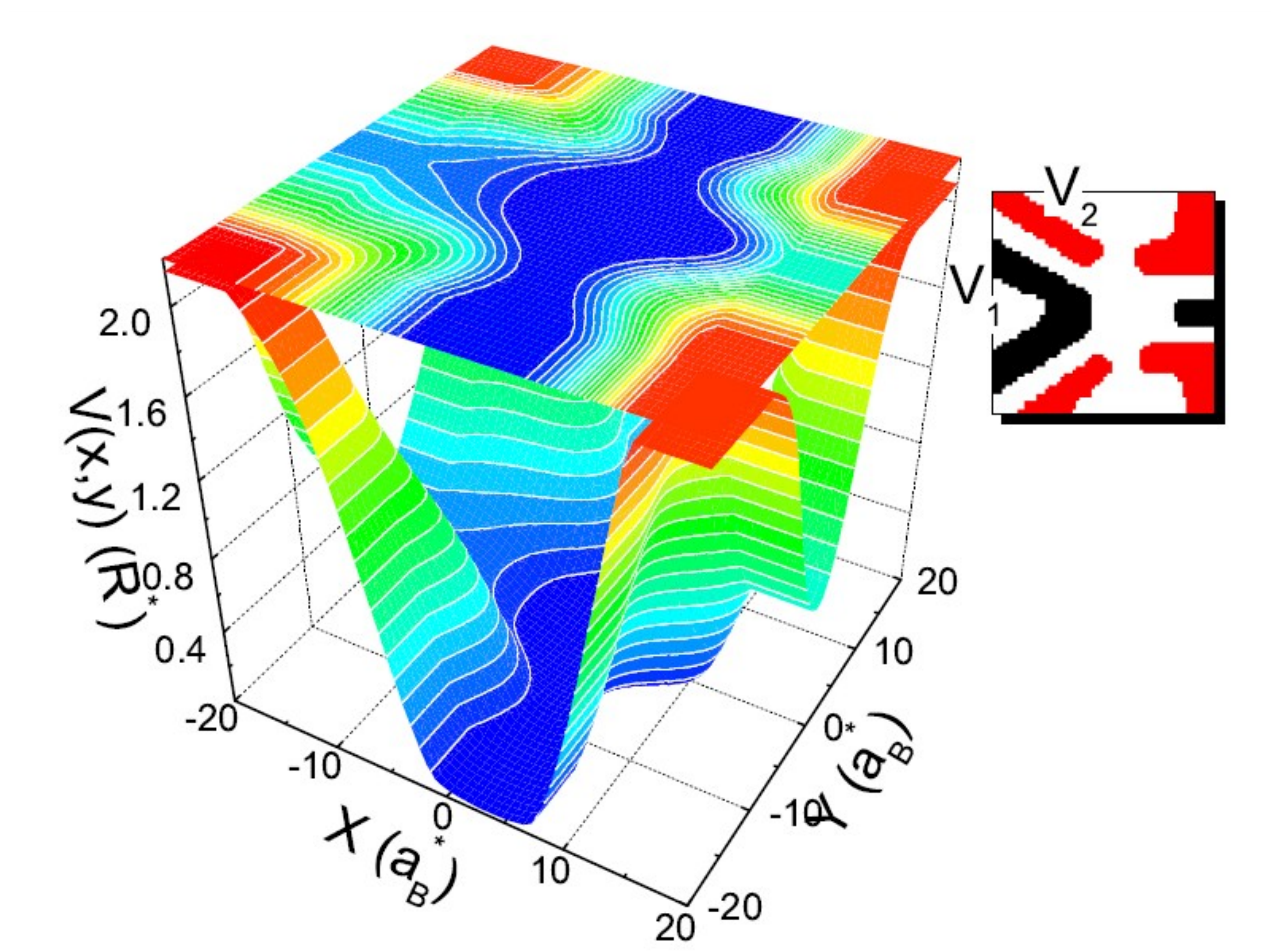}
\caption{ \label{figure:figure1} The self-consistent potential plotted
in 3D as a function of the lateral coordinates. The lengths are in
units of effective Bohr radius and energy is normalized with the
effective Rydberg energy. The inset depicts the sample geometry,
where S stands for source lead and D stands for the drain lead.
Coupling of the QD to the leads is manipulated by changing the
applied potential $V_2$.}}
\end{figure}
\section{Results and Discussion}
The calculation of the electrostatic potential considering real
sample geometries together with the electron-electron (e-e)
interaction is a challenging issue. Since such a calculation
cannot be done analytically for almost all the cases, usually
numerical techniques are deployed. It is clear that for "more than
a few" electron regime ($N>10$) exact diagonalization methods are
either impossible or very costly in terms of computational effort.
\begin{figure}
{\centering
\includegraphics[width=.7\linewidth]{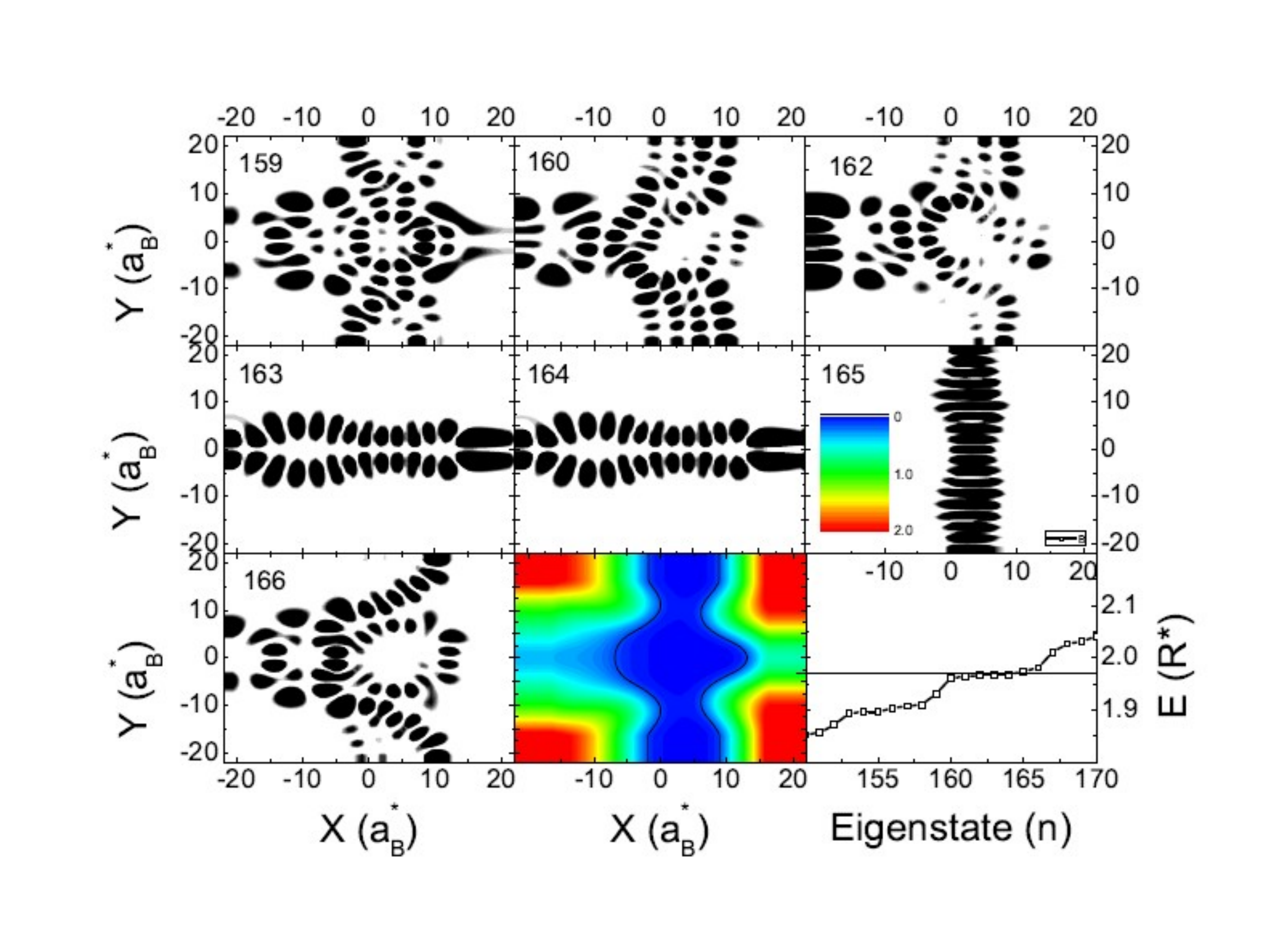}
\caption{ \label{figure:figure2} (a-g) Selected eigenfunctions residing
at the $E=1.96$ plateau (indicated by the horizontal line in i)
calculated for the color plotted potential (h), together with the
energy spectrum near $E_F$. The state $=160$ present a slight
asymmetry in coupling to source lead compared to the drain. The
asymmetric potential distribution with in the dot is visible.}}
\end{figure}
It is favorable to use a mean field approximation to describe the
(e-e) interactions, which is questionable in the "less than a few"
electron regime. The commonly used approach to determine the bare
confinement potential generated by the gates is the "frozen
charge" approximation~\cite{Davies95:4504}, which takes into
account properly the gate pattern and the effect of the spacer
between the gates and the 2DES. Since, it is not self-consistent
this approximation cannot account for the induced charges on the
metallic gates defining the QD. The effects resulting from the
induced charges and donor layer can be handled by solving the 3D
Poisson equation self-consistently. Almost a decade ago M. Stopa
introduced a very effective numerical scheme to describe the
electrostatics of such samples~\cite{Stopa96:selfcTF}, including
the e-e interactions either using a full Hartree, i.e. solving the
Poisson and Schr\"odinger equations self-consistently, or
considering Thomas-Fermi approximation (TFA). The
exchange-correlation interaction was accounted by a local density
approximation (LDA) using the density functional theory (DFT). It
was shown that the TFA is powerful enough to describe the
electrostatic potential even if the electrons are fully depleted
in some regions of the sample~\cite{Stopa96:selfcTF}.

Here, we stay in the TFA to calculate the electrostatic properties
of the real sample geometry using the algorithm developed by A.
Weichselbaum \textit{et.
al}~\cite{Andreas03:potential,Andreas:06}, which implements an
efficient grid relaxation technique to solve the 3D Poisson
equation. This approach was shown to be reliable to obtain the
potential profiles in the "more than a few" electron regime
considering QDs and quantum point contacts~\cite{Sefa08:prb}. The
next step in our calculation scheme is to obtain the single
particle energies and states, which we do same as described in the
previous section.

Figure.~\ref{figure:figure1} presents the calculated potential profile for
the sample geometry measured in Ref.~\cite{Heiblum05:abinter}. We
apply negative voltages to the gates shown in the inset. The upper
and lower two gates (denoted by red areas) are kept at the same
potential $V_2$, whereas the center gate (left black) and the
plunger gate (right black) are biased with a fixed voltage $V_1$.
Here, we consider a unit cell of 440x440 nm$^2$ spanned by 128x128
mesh matrix to calculate the self-consistent potential. The
surface potential is fixed to -0.75 V pinning the Fermi energy at
the mid gap. The 2DES is some 100 nm below the surface followed by
a thick GaAs layer. To achieve numerical convergence and satisfy
the open boundary conditions 3 mesh points of dielectric material
is assumed at all boundaries. In Figure.~\ref{figure:figure1} fixed
voltages of $V_1=-1.5$ V and $V_2=-2.2$ V are applied, the bulk
electron density is estimated to be $3\times10^{11}$ cm$^{-2}$
corresponding to $E_F\approx12.75$ meV, with the given density,
the number of electrons in the dot $N$ is similar to 200.
Figure.~\ref{figure:figure2} presents the calculated single particle wave
functions as a function of spatial coordinates, together with the
potential counter plot and the corresponding eigen energies versus
the state number. We show the states residing at the energy
\textit{plateau }, which lay in the close vicinity of $E_F$
(depicted by the horizontal solid line in Figure.~\ref{figure:figure2}i).
The states shown at the upper panel present the chaotic behavior,
whereas the first two states of
the mid panel are the non-propagating states. At $n=165$ a
resonant channel is observed, meanwhile the highest state shown
presents the chaotic behavior. These results show that,
qualitatively, transport through state 165 is much probable
compared to the others sitting at the same plateau. Although the
single particle energy eigenvalues are close to each other a
single channel is in charge of transport. At these gate voltages,
the QD is loosely defined as one can see that it is possible to
find an electron also at the left side of the actual QD. This
situation is changed by applying a higher negative potential to
the central and the plunger gates, $V_1=-2.0$ V. However, the QD
potential is not rotationally symmetric even if one neglects the
gates, since the center gate is geometrically different from the
plunger gate.
Now, we turn back our attention to the level width $\Gamma$ in a
qualitative manner. As we have discussed $\Gamma$ becomes
meaningful if one also considers both the source and drain leads.
\begin{figure}
{\centering
\includegraphics[width=.7\linewidth]{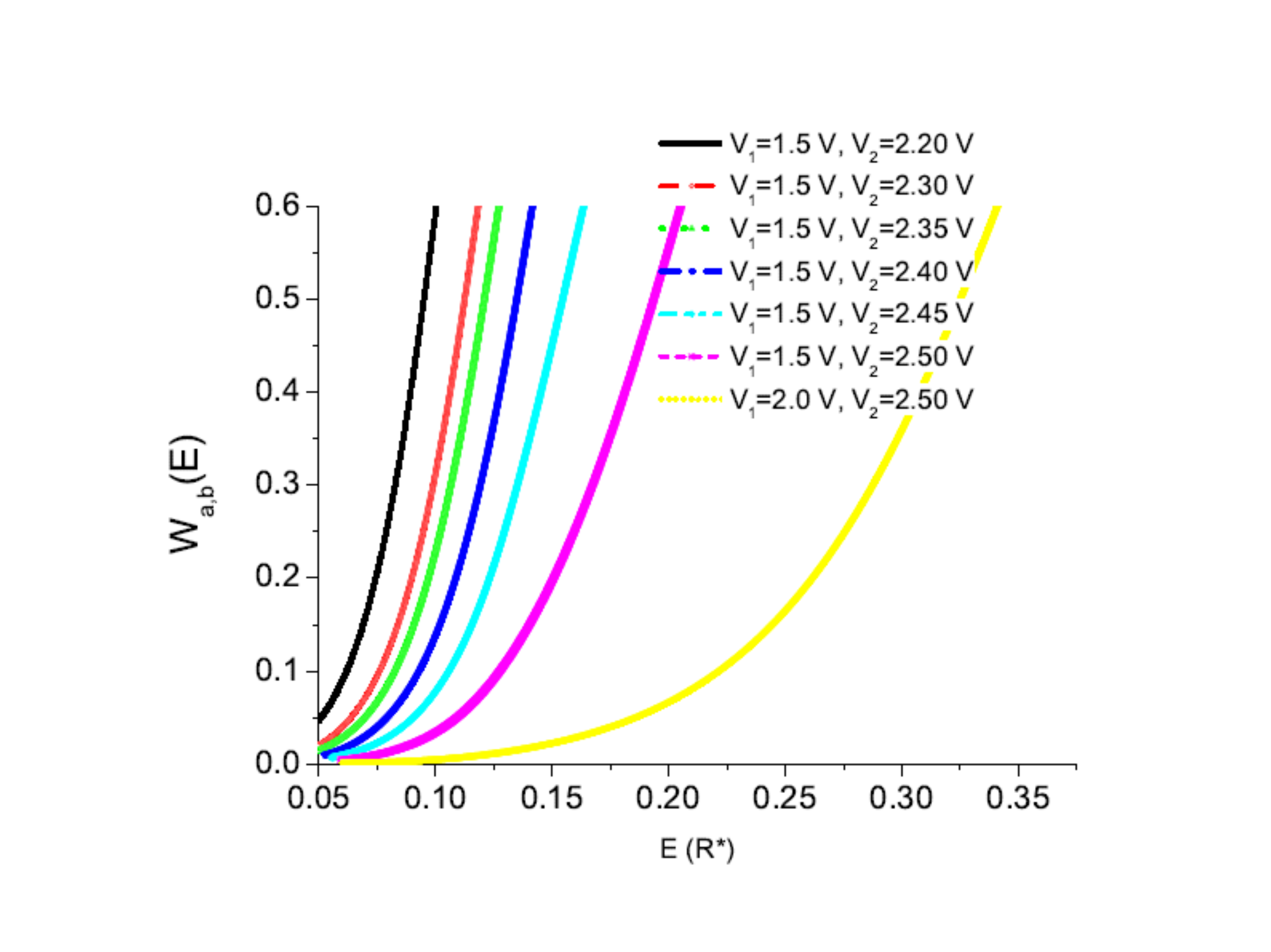}
\caption{ \label{figure:figure3} The transmission coefficients
calculated at the barrier. The turning points are obtained from
the self-consistent potential at the center of the barrier, where
the energy of the incoming wave cuts along.}}
\end{figure}
At this point it is useful to look at the transmission probability
$W_n(a,b)(E)$, in Fig.~\ref{figure:figure3} we show the this quantity as
a function of energy of the incoming wave calculated within the
WKB at various gate voltages, $V_1$, $V_2$. We see that, when the
upper and the lower gates are biased with small potentials, the
transmission increases linearly. This linearity changes if one
applies higher voltages to the barriers, however, for higher
energies the linearity is recovered. Such an observation leads us
to conclude that, essentially the
probability distribution determines the level widths, which may
become asymmetric considering transport at different energies.

To summarize, we have calculated the self-consistent electrostatic
potential exploiting the smooth variation of the bare potential
within the TFA. Next we obtained the single particle eigenstates
and energies considering a real sample geometry and crystal
structure. We found, similar to the Ref.~\cite{Imry00:towards}, that some single
particle levels bunch and present a energy plateau while changing
the state number. It was observed that, within these plateaus, not
all the states contribute to the transport since the overlap of
the dot wave functions and lead wave functions simply vanish. More
interestingly, we found that at intermediately high energies, the
wave functions are coupled to at least one of the leads much
stronger than the ones in their close energy vicinity. This
result, we believe, supports the phenomenological models, which
attribute the abrupt change of the phase lapses to
electron-electron interactions.
\ack
We would like to thank Jan von Delft for introducing us the ``phase lapses'' problem and motivating us to perform numerical calculations. Moty Heiblum is also acknowledged for his enlightening discussions. This work is partially supported by TUBiTAK, under grant no:109T083.
\section*{References}


\bibliographystyle{unsrt}
\end{document}